# Interplay of Landau quantization and interminivalley scatterings in a weakly coupled moiré superlattice

Yalong Yuan[1,2,†], Le Liu[1,2,†], Jundong Zhu[1,2], Jingwei Dong[1,2], Yanbang Chu[1,2], Fanfan Wu[1,2], Luojun Du[1,2], Kenji Watanabe[4], Takashi Taniguchi[5], Dongxia Shi[1,2,3], Guangyu Zhang[1,2,3*] & Wei Yang[1,2,3*]

[1] *Beijing National Laboratory for Condensed Matter Physics and Institute of Physics, Chinese Academy of Sciences, Beijing 100190, China*
[2] *School of Physical Sciences, University of Chinese Academy of Sciences, Beijing, 100190, China*
[3] *Songshan Lake Materials Laboratory, Dongguan 523808, China*
[4] *Research Center for Electronic and Optical Materials, National Institute for Materials Science, 1-1 Namiki, Tsukuba 305-0044, Japan*
[5] *Research Center for Materials Nanoarchitectonics, National Institute for Materials Science, 1-1 Namiki, Tsukuba 305-0044, Japan*

[†] These authors contributed equally to this work.
* Corresponding authors. Email: gyzhang@iphy.ac.cn; wei.yang@iphy.ac.cn

**Abstract**

Double layer quantum systems are promising platforms for realizing novel quantum phases. Here, we report a study of quantum oscillations (QOs) in a weakly coupled double layer system, composed of a large angle twisted double bilayer graphene (TDBG). We observe two different QOs at low temperature, one with a periodicity in carrier density ($n$), i.e. Shubnikov–de Haas oscillation (SdHO) due to Landau quantization, and the other one in displacement field ($D$), resulting a grid pattern. We quantify the interlayer coupling strength by measuring the interlayer capacitance from the grid pattern with a capacitance model, revealing an electron-hole asymmetry. At high temperature when SdHO are thermal smeared, we observe resistance peaks when LLs from two minivalleys in the moiré Brillion zone are aligned, regardless of carrier density; eventually, it results in a two-fold increase of oscillating frequency in $D$, serving as a smoking-gun evidence of the magneto-intersubband oscillations (MISO) in a double layer system. The temperature dependence of MISO suggests electron-electron interaction between two minivalleys play a crucial rule in the scattering, and the scattering times obtained from MISO thermal damping are found to be correlated with the interlayer coupling strength. Our study reveals an intriguing interplay among Landau quantization, moiré band structure, and scatterings.

**Main Text**

Double layer quantum systems are ideal platforms for realizing novel quantum phases[1–9], thanks to the rich interplay among charge, spin, and layer degree of freedom. Recently, twisted graphene multilayers have been demonstrated as a promising paradigm for developing novel states of matter[10–18]. A key ingredient lies in the twist angle, which determines not only the lateral size of moiré superlattice in real space, i.e. moiré Brillouin zone in reciprocal space, but also the coupling between the layers. At some optimal angle, strong interlayer coupling will lead to flat moiré bands with the hybridized electronic states indistinguishable between the two layers, which favors correlated states[19–23]. At a larger angle, in the so-called weak coupling regime, the moiré superlattice reduces in size and the moiré bands are usually dispersive with a reduced fermi velocity, and importantly the wave function of minivalleys is layer polarized[24–28]. Essentially, the interlayer coupling defines the electronic correlation between two layers, and to tune this coupling is crucial for realizing collective quantum phases in various double layer systems. The ability to quantify the coupling in double layer systems is important yet rarely explored.

Here, we focus on a weak coupled double layer moiré system, which is composed of a large angle twisted double bilayer graphene (TDBG), by measuring the quantum oscillations (QOs) due to Landau quantization. The QOs, in principle, are determined by the combined band structure of the two layers, and thus could be serve as a probe for the interlayer coupling. Displacement field ($D$) is a powerful knob in tuning electronic structure in TDBG. In the strong coupling regime, it helps to reveal spin polarized correlated insulators and valley polarized insulators at different $D$[12–14,29–31], and even leads to insulating QOs in band inversion condition[32]. In the weak coupling regime, $D$ drives a charge redistribution between the two layers, or equivalently between two minivalleys[24,26,27]. The scatterings between layers are then reflected on the interminivalley scatterings, as shown in Fig. 1(b) and 1(e). For a twist angle of around 2°, due to the small size of the moiré Brillion zone (Fig. 1(c)), scatterings between minivalleys could be significant[25,27]. In the following, we will first quantify the interlayer coupling strength from the periodicity of the QOs at low temperature in a capacitance model of $D$-field tuned SdHO from two minivalleys, yielding an electron-hole asymmetry. Then, we reveal a two-fold increase of oscillating frequency in $D$ for the QOs at elevated temperature when SdHO are thermal smeared, and we attribute them as the enhanced interminivalley scatterings when the LLs at $k$ and $k$' minivalleys are aligned. Lastly, we demonstrate an intimate connection between interlayer coupling strength and the interminivalley scatterings, by quantifying the scattering lifetime from the high temperature oscillating magnetoresistance.

We fabricate ultraclean TDBG devices by the cut and stack methods[33–35]. In the following, we focus on device D1 with a twist angle of 2.25° that reveals high-quality oscillation data with low resistance of few tens of ohms. The device is in a dual gate geometry, allowing an independent tuning of total carrier density $n$ and displacement field $D$, i.e. $n = (C_t V_t + C_b V_b)/e$ and $D/\varepsilon_0 = (C_t V_t - C_b V_b)/2/\varepsilon_0$, where $C_b$ ($C_t$) is the geometrical capacitance per area for bottom (top) gate, $e$ is the electron charge, and $\varepsilon_0$ is vacuum permittivity. Fig. 2(a) is a color mapping of measured longitudinal resistance $R_{xx}$ as a function of $n$ and $D/\varepsilon_0$ at $B = 2$ T. The oscillations due to Landau quantization are vividly revealed in grid pattern, indicated by the alternative changes of the color between blue and white. The oscillation data are plotted in Fig. 2(b), where the resistance is minimum at LL filling factor $v_{LL} = 8N$ for $D/\varepsilon_0 = 0$, and at $v_{LL} = 8(N+1/2)$ for $D/\varepsilon_0 = -0.04$ V/nm, with $N$ a nonzero integer. The data is of high quality, with a dip value of ~15 Ω and oscillation amplitude of ~60 Ω at around $v_{LL} = -50$. $v_{LL} = n\Phi_0/B$ is the LL filling factor and $\Phi_0$ is magnetic flux quantum.

To begin with, we consider Landau level crossings of the two minivalleys that are approximated as two layer-polarized parabolic energy bands (Fig. 2(d)). They are quantized into a series of Landau levels (LLs) at a given $B$, described by

$$E_{t/b} = (N_{t/b} + 1/2)\ \hbar\omega \mp e\Delta V/2. \quad (1)$$

Here $t/b$ corresponds to the top/bottom layer, $N$ is the Landau level index, $\hbar$ is the reduced Planck constant, $\omega = eB/m^*$ is the cyclotron frequency, and $\Delta V$ is the interlayer potential difference that depends on $D$. With the change of $\Delta V$, the two sets of LLs will intersect each other as shown in Fig. 2(e). By taking $m^* = 0.08\ m_e$ and LLs broadening of 0.35 $\hbar\omega$, the calculated Landau level DOS mapping (details are shown in Methods) is shown in Fig. 2(c). The calculations agree well with the experimental data in Fig. 2(a), indicating a linear relationship between $\Delta V$ and $D$ in the moiré valence band.

Next, we quantitatively analyze the QOs to reveal the intimate link between $\Delta V$ and $D$. The TDBG device follows a capacitive model[36–38] (Fig. 1(e) and details in supporting information), and it gives

$$\Delta V = (D + e\Delta n/2)/C_m, \quad (2)$$

where $C_m$ is measured interlayer capacitance and $\Delta n$ is the carrier density difference between the top and bottom bilayer graphene. The magnitude of interlayer capacitance $C_m$ reflects the degree of layer polarization[39,40], and the larger the $C_m$ the stronger the coupling between two twisted layers. Experimentally, we can extract the interlayer capacitance $C_m$ according to the equation (2). At the intersection point of two sets of LLs in Fig. 2(a) and 2(c), one could obtain $\Delta n_{e/h} = \pm(N_t - N_b) \times 4eB/h$ and $\Delta V_{e/h} = \mp(N_t - N_b) \times \hbar B/m^*$, where the subscript $e/h$ represents the electron/hole and $m^*$ is 0.08 $m_e$ for holes and 0.06 $m_e$ for electrons extracted from the thermal damping effect of QOs at $D/\varepsilon_0 = 0$. The obtained $C_m$ is plotted against $v_{LL}$ in Fig. 2(f). For valence band, $C_m \sim 6.3$ μF/cm$^2$ is constant, regardless of $v_{LL}$; this suggests the valance band is a well-defined decoupled system[36,40]. For conduction band, on the contrary, it shows a strong carrier density dependence. $C_m$ reaches up to 19 μF/cm$^2$ near the CNP, and it decreases and tends to saturate at a small value of ~7 μF/cm$^2$ at high density in Fig. 2(c). Such a large capacitance indicates a strong coupling in the conduction band, and its strong doping dependence suggests the interlayer coupling strength is highly sensitive to the change of fermi surface. Note that similar results are reproduced in device D2 (Supplementary Fig. 16). It is also worth noted that the observed electron-hole asymmetry in capacitance model matches well with our continuum model calculations, where the degree of layer polarization for holes is higher than that for electrons (Supplementary Fig. 4).

To investigate the scatterings in the weakly coupled double layer system, we measure the QOs at high temperature. Fig. 3(a) shows a color mapping of longitudinal resistance $R_{xx}$ ($n$, $D/\varepsilon_0$) at $B$ = 2 T and $T$ = 20 K, and it reveals a striped pattern, unlike the previous grid pattern at low temperature. The disappearance of oscillating $R_{xx}(n)$ at a fixed $D$ indicates that conventional SdHOs are thermal smeared. The striped patterns depend only by $D$, irrespective of total carrier density $n$; in other words, the strip patterns are determined only by carrier density difference between top and bottom layer ($k$ and $k'$ minivalley). The observation agrees perfectly with resonant condition when LLs from two minivalleys are aligned, and are the main features of the magneto-intersubband oscillations (MISO)[24,41–46]. Take the dashed line in Fig. 3(a) for example, comparisons of $R_{xx}(D/\varepsilon_0)$ at different $T$ are shown in Fig. 3(c). The transition from SdHO to MISO is captured in the doubling of $D$-periodicity as $T$ is increased from 1.8 K to ≥10 K. Such transition is also vividly captured in the temperature evolution of the peak width and peak amplitude at $D/\varepsilon_0$ = 0 in Fig. 3(d); it decreases rapidly from ~26 mV/nm at $T$ < 4K to ~16 mV/nm at 10 K, and then it barely changes at $T$ ≥10 K. Note that, by considering the spacing between peaks at $T$ ≥10 K as $\hbar\omega$, a peak width of ~16 mV/nm is corresponding to an energy linewidth of ~1.85 meV. In addition, the high temperature MISO are found emanating from $D/\varepsilon_0$ = 0 at $B$ = 0 T in Fig. 3(b), which resembles the conventional Landau fan diagram when $n$ is varied. The resemblance indicates that the carrier density difference between two minivalleys are linearly generated by $D$. And most importantly, the resistance peak at $D/\varepsilon_0$ = 0 indicates additional scattering processes for the resonant condition, as shown in the schematics of Fig. 3(e), which is exactly the predicted interminivalley scatterings in the MISO model. Similar observations are obtained for electrons in Supplementary Note 6 and 7.

To further establish a close connection between the interlayer coupling obtained at low temperature and the interminivalley scatterings at high temperature, we measure the resistance as a function of $B$ at high temperature. At a fixed carrier density of -2.16 × $10^{12}$ /cm$^2$ in the valence band, the temperature ($T$) dependent QOs are plotted against $1/B$ for $D/\varepsilon_0$ = 0 and -0.08 V/nm in Fig. 4(a) and 4(b), respectively. For $D/\varepsilon_0$ = -0.08 V/nm, it evolves from a beating pattern at low $T$ to a single frequency oscillation at high $T$. The changes are better revealed in the fast Fourier transform (FFT) spectrum in the inset of Fig. 4(b), where the oscillation frequency ($B_f$) changes from three peaks at $T$ = 1.8 K to one peak at $T$ = 15 K. By contrast, for $D/\varepsilon_0$ = 0 V/nm in Fig. 4(a), $B_f$ ~ 11 T keeps unchanged while its amplitude quickly fades, as $T$ is increased. Note that the frequency $B_f$ is linearly correlated with $n$, defined as $B_f = nh/ge$. Considering the degeneracy $g$ = 4 for each minivalley, the total degeneracy is 8 at $D/\varepsilon_0$ = 0 V/nm and $B_f$ ~11 T is corresponding to $n$ = 2.13 × $10^{12}$ /cm$^2$ that matches well with the gate induced carrier density, which is then assigned as SdHO. At nonzero $D/\varepsilon_0$ = -0.08 V/nm, minivalleys degeneracy is lifted, and the three oscillation modes at $B_f$ ~4, 8.2, and 12.2 T in Fig. 4(b) are corresponding to $n$ = 0.387, 0.794, and 1.181 × $10^{12}$ /cm$^2$, respectively. The two high frequency modes in Fig. 4(b) share the same $T$ dependence as the one in Fig. 4(a), and are attributed to the SdHO in two minivalley $k$ and $k'$. What's more, the lowest frequency mode at $B_f = 4$ T is equal to the difference $B_f(k) - B_f(k')$; in other words, the slow oscillation mode is determined by minivalley carrier density difference $\Delta n$. The observation of the low frequency mode could persist at high $T$ resembles the high temperature QOs in weak coupled TBG, is another characteristic of MISO[24,41–46] in TDBG.

The details of the scattering lifetime are quantitatively revealed in the thermal damping of the QOs. The thermal damping of SdHO is described by Lifshitz-Kosevich (L-K) formula, i.e. $R(T) \sim (2\pi^2 k_B T/\hbar\omega)/\sinh(2\pi^2 k_B T/\hbar\omega)$. As shown in Fig. 4(c), the amplitude of SdHO is barely visible at $T = 10$ K, and similar $T$ dependence for $B_1$ and $B_2$ indicate identical effective mass for the minivalleys. By contrast, MISO observed in Fig. 4(c) are thermal damped with a slow temperature dependence described as $\exp(-\gamma T^2)$ (Fig. 4(c)). In a clean limit ($\omega\tau \gg 1$, $\tau$ is the total scattering time)[41,42,44,47,48] with the assumption of identical scattering time for two minivalley, MISO in TDBG could be described as:

$$\Delta\rho = 2(\tau/\tau_{\text{inter}})\rho_0 \exp(-2\pi/(\omega\tau_q)) \cos(2\pi(E_t-E_b)/\hbar\omega). \qquad (3)$$

Here $\rho_0$ is the Drude resistivity; $\tau$ and $\tau_{\text{inter}}$ are the total and interminivalley scattering time, respectively; $\exp(-2\pi/(\omega\tau_q))$ is the Dingle factor, with $\tau_q$ the quantum scattering time. The observed decay of $\exp(-\gamma T^2)$ indicates a scattering rate of $\sim T^2$, which further suggests an important role played by electron-electron interaction[44]. We attribute the prominent quadratic law to the enhanced Umklapp electron scatterings in moiré system, which also agrees with the quadratic power law of zero-field resistivity at low temperatures in Supplementary Fig.14 when phonon contributions are negligible in TDBG[49,50]. At a fixed $T = 18$ K, the $\log(\Delta\rho)$ vs $1/B$ gives quantum scattering time $\tau_q \sim 1.04$ ps, which is of the same order of magnitude compared to the mean free time; by changing the $T$, we could obtain $\tau_q$ vs $T$, see Supplementary Fig. 12. It is worth mentioning that the low resistance, tens of ohm in device D1, as well repeatable results in device D2, enables us to reveal the intrinsic MISO and to rule out other extrinsic impurities or phonon scatterings mechanisms, distinct from previous report in TDBG. In addition, we can quantitively extract the ratio of $\tau/\tau_{\text{inter}}$ by fitting the high temperature magnetoresistance curve with the equation (3). Fig.4(d) shows the obtained $\tau/\tau_{\text{inter}}$ for both the electrons and holes. For holes, the ratio is small, ranging from 0.048 to 0.2, indicating a small intervalley scatting rate $\Gamma \sim 1/\tau_{\text{inter}}$; while, for electrons, it is much larger, ranging from ~0.4 to 0.8, indicating a comparable contribution between intravalley and intervalley scatterings. For instance, at a fixed $n = \pm 2.0 \times 10^{12}$ /cm$^2$ and $T = 18$ K (Supplementary Fig. 11 and 13), one could obtain an intervalley scatter time of ~120 ps and ~6 ps for holes and electrons, respectively. It is worthy to note that the obtained $\tau/\tau_{\text{inter}}$ also agrees well with the electron-hole asymmetric interlayer coupling from $C_m$ in Fig. 2(c), thus establishing an intimate connection between intervalley scatterings and interlayer couplings.

In conclusion, we systematically investigate the interlayer coupling and interminivalley scatterings from quantum oscillations in a weakly coupled moiré superlattice of TDBG. The results show that the valance band is a well-defined decoupled system, whereas the interlayer coupling strength of conduction band highly depends on the doping levels, indicating strong electron-hole asymmetry. In addition, we demonstrate strong interminivalley scattering in this system according to the observed MISO for both electrons and holes. The MISO can be effectively tuned by $D$, linearly via $\Delta E(D) \sim D$; the temperature dependence of MISO suggests electron-electron interactions and interlayer couplings play a crucial rule in the scattering. Being a highly tunable moiré system, TDBG provides an excellent platform to investigate the interplay among Landau quantization, moiré band structure, and scatterings.

**Methods**

**1. Device fabrication and characterization.**

The twisted double bilayer graphene (TDBG) devices are fabricated by the cut and stack technique[33]. Bilayer graphene, few-layer graphite (FLG), and hexagonal boron nitride (h-BN) (15–40 nm thick) were firstly exfoliated on $SiO_2$. As shown in Supplementary Fig. 1, the layer number of graphene is identified by the reflection contrast from the optical microscope images[51] as well as the Raman spectrum[52]. Then we used an atomic force microscope (AFM) needle to cut a large bilayer graphene flake into two pieces. Next, we used poly (bisphenol A carbonate) (PC) supported by polydimethylsiloxane (PDMS) on a glass slide to pick up the top h-BN, two pieces of bilayer graphite with a twist angle of 60° + $\theta$, bottom h-BN, and bottom graphite gate in turn.

The fabrication of the metal top gate and electrodes followed a standard electron-beam lithography process and electron-beam metal evaporation. The devices were designed as a Hall bar structure and shaped by traditional reactive ion etching with a $CHF_3$ and $O_2$ gas mixture. Finally, Cr/Au electrodes are evaporated on silicon wafer as 1D edge contacts to connect with Hall bars. The combination of two gate voltages allows independent tuning of the carrier density $n$ and $D$, i.e. $n = (C_t V_t + C_b V_b)/e$ and $D = (C_t V_t - C_b V_b)/2$.

The transport behaviors are measured at the base temperature (~1.8 K) of our helium-4 cryostat. The twisted angle of the TDBG devices are determined by the carrier density difference between CNP and gap opening at full filling, i.e. $n_s = 4/A \approx 8\theta/(\sqrt{3}a^2)$, where $A$ is the area of a moiré unit cell, $\theta$ is twisted angle, and $a$ is the lattice constant of graphene.

**2. Continuum model calculations.**

The band structure of TDBG is calculated by continuum model[19,53]. In calculation, we use the parameters from reference[27,53]

$$(\gamma_0, \gamma_1, \gamma_3, \gamma_4, w_{AA}, w_{AB}) = (3100, 400, 320, 44, 100, 100) \text{ meV}$$

Here $\gamma_0$ is the intralayer nearest-neighbor hopping term; $\gamma_1$ is the interlayer hopping term in AB-stacked bilayer graphene, from bottom layer A sublattice to top layer B sublattice; $\gamma_3$ is the interlayer hopping term from bottom layer B sublattice to top layer A sublattice; γ4 is the interlayer hopping term from bottom layer A(B) sublattice to top layer A(B) sublattice. γ3 and γ4 will bring electron-hole asymmetry in twisted double bilayer graphene systems. The electric displacement field changes the on-site energy of every layer in twisted graphene systems, described by the following matrix,

$$V = \begin{pmatrix} \frac{U}{2}\hat{1} & 0 & 0 & 0 \\ 0 & \frac{U}{6}\hat{1} & 0 & 0 \\ 0 & 0 & -\frac{U}{6}\hat{1} & 0 \\ 0 & 0 & 0 & -\frac{U}{2}\hat{1} \end{pmatrix}$$

where $U$ is electrostatic interlayer potential energy, $\hat{1}$ is a 2×2 unit matrix.

## 3. Landau levels spectrum calculations.

The band structure of $k$ and $k'$ minivalley is approximately equivalent to two parabolic bands localized in top and bottom layers, respectively. Now we focus on the case of hole-type valance band. The Landau levels energy spectrum can be described by the following formulas.

$$E_{tn} = (n + 1/2)h\omega - eV_0/2$$
$$E_{bn} = (n + 1/2)h\omega + eV_0/2$$
$$\omega = eB/m^*$$

Next, we calculate the density of states (DOS) of Landau levels based on the Gauss distribution by considering the disorder induced Landau levels broaden.

$$D_n(E) = gn_L \sqrt{\frac{2}{\pi \Gamma^2}} \exp\left(-2\frac{(E-E_n)^2}{\Gamma^2}\right)$$

Here $\Gamma$ corresponds to the broaden of Landau levels, and we use an estimation of $\Gamma/hw = 0.35$ to fit our experimental data. $g$ is the degeneracy of energy band and $n_L$ is the carrier density to full fill a Landau level. At a fixed interlayer voltage $V$, DOS can be expressed as

$$D_V(E) = \sum_n (D_n^{up}(E) + D_n^{down}(E))$$

The filling factor $v$ is calculated by integrating DOS with $E$.

$$v = n/n_L = \frac{1}{n_L} \int D_U(E) dE$$

**Acknowledgements**

We acknowledge supports from the National Key Research and Development Program (Grant No. 2020YFA0309600), National Science Foundation of China (NSFC, Grant Nos. 12074413, 61888102, 11834017), the Strategic Priority Research Program of CAS (Grant Nos. XDB33000000) and the Key-Area Research and Development Program of Guangdong Province (Grant No. 2020B0101340001). K.W. and T.T. acknowledge support from the JSPS KAKENHI (Grant Numbers 20H00354, 21H05233 and 23H02052) and World Premier International Research Center Initiative (WPI), MEXT, Japan.

## Author contributions

W.Y. and G.Z. supervised the project. W.Y., L.L. designed the experiments. Y.Y., L.L. fabricated the devices. Y.L., L.L. performed the magneto-transport measurement. L.L. performed the calculations. K.W. and T.T. provided hexagonal boron nitride crystals. Y.Y., L.L., G.Z. and W.Y. analyzed the data. W.Y., Y.Y., L.L. wrote the paper with the input from all the authors.

## Data availability

The data that support the findings of this study are available from the corresponding authors upon reasonable request.

## Competing interests

The authors declare no competing interests.

## Additional information

Supplementary information is provided online.

## References

1. Eisenstein, J. P., Boebinger, G. S., Pfeiffer, L. N., West, K. W. & He, S. New fractional quantum Hall state in double-layer two-dimensional electron systems. *Phys. Rev. Lett.* **68**, 1383–1386 (1992).
2. Suen, Y. W., Engel, L. W., Santos, M. B., Shayegan, M. & Tsui, D. C. Observation of a $\nu=1/2$ fractional quantum Hall state in a double-layer electron system. *Phys. Rev. Lett.* **68**, 1379–1382 (1992).
3. Eisenstein, J. P. & MacDonald, A. H. Bose–Einstein condensation of excitons in bilayer electron systems. *Nature* **432**, 691–694 (2004).
4. Fogler, M. M., Butov, L. V. & Novoselov, K. S. High-temperature superfluidity with indirect excitons in van der Waals heterostructures. *Nat. Commun.* **5**, 4555 (2014).
5. Du, L. *et al.* Evidence for a topological excitonic insulator in InAs/GaSb bilayers. *Nat. Commun.* **8**, 1971 (2017).
6. Li, J. I. A., Taniguchi, T., Watanabe, K., Hone, J. & Dean, C. R. Excitonic superfluid phase in double bilayer graphene. *Nat. Phys.* **13**, 751–755 (2017).
7. Liu, X., Watanabe, K., Taniguchi, T., Halperin, B. I. & Kim, P. Quantum Hall drag of exciton condensate in graphene. *Nat. Phys.* **13**, 746–750 (2017).
8. Ma, L. *et al.* Strongly correlated excitonic insulator in atomic double layers. *Nature* **598**, 585–589

(2021).

9. Liu, X. *et al.* Crossover between strongly coupled and weakly coupled exciton superfluids. *Science* **375**, 205–209 (2022).

10. Cao, Y. *et al.* Correlated insulator behaviour at half-filling in magic-angle graphene superlattices. *Nature* **556**, 80–84 (2018).

11. Cao, Y. *et al.* Unconventional superconductivity in magic-angle graphene superlattices. *Nature* **556**, 43–50 (2018).

12. Cao, Y. *et al.* Tunable correlated states and spin-polarized phases in twisted bilayer–bilayer graphene. *Nature* **583**, 215–220 (2020).

13. Liu, X. *et al.* Tunable spin-polarized correlated states in twisted double bilayer graphene. *Nature* **583**, 221–225 (2020).

14. Shen, C. *et al.* Correlated states in twisted double bilayer graphene. *Nat. Phys.* **16**, 520–525 (2020).

15. Serlin, M. *et al.* Intrinsic quantized anomalous Hall effect in a moiré heterostructure. *Science* **367**, 900–903 (2020).

16. Xu, S. *et al.* Tunable van Hove singularities and correlated states in twisted monolayer–bilayer graphene. *Nat. Phys.* **17**, 619–626 (2021).

17. Park, J. M., Cao, Y., Watanabe, K., Taniguchi, T. & Jarillo-Herrero, P. Tunable strongly coupled superconductivity in magic-angle twisted trilayer graphene. *Nature* **590**, 249–255 (2021).

18. Zhang, Y. *et al.* Promotion of superconductivity in magic-angle graphene multilayers. *Science* **377**, 1538–1543 (2022).

19. Bistritzer, R. & MacDonald, A. H. Moiré bands in twisted double-layer graphene. *Proc. Natl. Acad. Sci.* **108**, 12233–12237 (2011).

20. Po, H. C., Zou, L., Vishwanath, A. & Senthil, T. Origin of Mott Insulating Behavior and Superconductivity in Twisted Bilayer Graphene. *Phys. Rev. X* **8**, 031089 (2018).

21. Tarnopolsky, G., Kruchkov, A. J. & Vishwanath, A. Origin of Magic Angles in Twisted Bilayer Graphene. *Phys. Rev. Lett.* **122**, 106405 (2019).

22. Lee, J. Y. *et al.* Theory of correlated insulating behaviour and spin-triplet superconductivity in twisted double bilayer graphene. *Nat. Commun.* **10**, 5333 (2019).

23. Lisi, S. *et al.* Observation of flat bands in twisted bilayer graphene. *Nat. Phys.* **17**, 189–193 (2021).

24. de Vries, F. K. *et al.* Combined Minivalley and Layer Control in Twisted Double Bilayer Graphene. *Phys. Rev. Lett.* **125**, 176801 (2020).

25. Phinney, I. Y. *et al.* Strong Interminivalley Scattering in Twisted Bilayer Graphene Revealed by High-Temperature Magneto-Oscillations. *Phys. Rev. Lett.* **127**, 056802 (2021).

26. Rickhaus, P. *et al.* Correlated electron-hole state in twisted double-bilayer graphene. *Science* **373**, 1257–1260 (2021).

27. Tomić, P. *et al.* Scattering between Minivalleys in Twisted Double Bilayer Graphene. *Phys. Rev. Lett.* **128**, 057702 (2022).

28. Kim, Y., Moon, P., Watanabe, K., Taniguchi, T. & Smet, J. H. Odd Integer Quantum Hall States
9

with Interlayer Coherence in Twisted Bilayer Graphene. *Nano Lett.* **21**, 4249–4254 (2021).

29. He, M. *et al.* Symmetry breaking in twisted double bilayer graphene. *Nat. Phys.* **17**, 26–30 (2021).
30. Liu, L. *et al.* Isospin competitions and valley polarized correlated insulators in twisted double bilayer graphene. *Nat. Commun.* **13**, 3292 (2022).
31. Liu, L. *et al.* Observation of First-Order Quantum Phase Transitions and Ferromagnetism in Twisted Double Bilayer Graphene. *Phys. Rev. X* **13**, 031015 (2023).
32. Liu, L. *et al.* Quantum oscillations in field-induced correlated insulators of a moiré superlattice. *Sci. Bull.* **68**, 1127–1133 (2023).
33. Kim, K. *et al.* van der Waals Heterostructures with High Accuracy Rotational Alignment. *Nano Lett.* **16**, 1989–1995 (2016).
34. Pizzocchero, F. *et al.* The hot pick-up technique for batch assembly of van der Waals heterostructures. *Nat. Commun.* **7**, 11894 (2016).
35. Purdie, D. G. *et al.* Cleaning interfaces in layered materials heterostructures. *Nat. Commun.* **9**, 5387 (2018).
36. Sanchez-Yamagishi, J. D. *et al.* Quantum Hall Effect, Screening, and Layer-Polarized Insulating States in Twisted Bilayer Graphene. *Phys. Rev. Lett.* **108**, 076601 (2012).
37. Cheng, B. *et al.* Gate-Tunable Landau Level Filling and Spectroscopy in Coupled Massive and Massless Electron Systems. *Phys. Rev. Lett.* **117**, 026601 (2016).
38. Mreńca-Kolasińska, A. *et al.* Quantum capacitive coupling between large-angle twisted graphene layers. *2D Mater.* **9**, 025013 (2022).
39. Fang, J., Vandenberghe, W. G. & Fischetti, M. V. Microscopic dielectric permittivities of graphene nanoribbons and graphene. *Phys. Rev. B* **94**, 045318 (2016).
40. Rickhaus, P. *et al.* The electronic thickness of graphene. *Sci. Adv.* **6**, eaay8409 (2020).
41. Bykov, A. A. Nonlinear magnetotransport in a high-mobility quasi-two-dimensional electron system. *JETP Lett.* **88**, 64–68 (2008).
42. Mamani, N. C. *et al.* Classical and quantum magnetoresistance in a two-subband electron system. *Phys. Rev. B* **80**, 085304 (2009).
43. Wiedmann, S., Gusev, G. M., Raichev, O. E., Bakarov, A. K. & Portal, J. C. Thermally activated intersubband scattering and oscillating magnetoresistance in quantum wells. *Phys. Rev. B* **82**, 165333 (2010).
44. Dmitriev, I. A., Mirlin, A. D., Polyakov, D. G. & Zudov, M. A. Nonequilibrium phenomena in high Landau levels. *Rev. Mod. Phys.* **84**, 1709–1763 (2012).
45. Bykov, A. A. *et al.* Beats of Quantum Oscillations of the Resistance in Two-Subband Electron Systems in Tilted Magnetic Fields. *JETP Lett.* **109**, 400–405 (2019).
46. Minkov, G. M. *et al.* Magneto-intersubband oscillations in two-dimensional systems with an energy spectrum split due to spin-orbit interaction. *Phys. Rev. B* **101**, 245303 (2020).
47. Raikh, M. E. & Shahbazyan, T. V. Magnetointersubband oscillations of conductivity in a two-dimensional electronic system. *Phys. Rev. B* **49**, 5531–5540 (1994).

48. Mamani, N. C., Gusev, G. M., Raichev, O. E., Lamas, T. E. & Bakarov, A. K. Nonlinear transport and oscillating magnetoresistance in double quantum wells. *Phys. Rev. B* **80**, 075308 (2009).

49. Li, X., Wu, F. & Das Sarma, S. Phonon scattering induced carrier resistivity in twisted double-bilayer graphene. *Phys. Rev. B* **101**, 245436 (2020).

50. Chu, Y. *et al.* Temperature-linear resistivity in twisted double bilayer graphene. *Phys. Rev. B* **106**, 035107 (2022).

51. Li, H. *et al.* Rapid and Reliable Thickness Identification of Two-Dimensional Nanosheets Using Optical Microscopy. *ACS Nano* **7**, 10344–10353 (2013).

52. Graf, D. *et al.* Spatially Resolved Raman Spectroscopy of Single- and Few-Layer Graphene. *Nano Lett.* **7**, 238–242 (2007).

53. Koshino, M. Band structure and topological properties of twisted double bilayer graphene. *Phys. Rev. B* **99**, 235406 (2019).

**Figure & Figure captions**

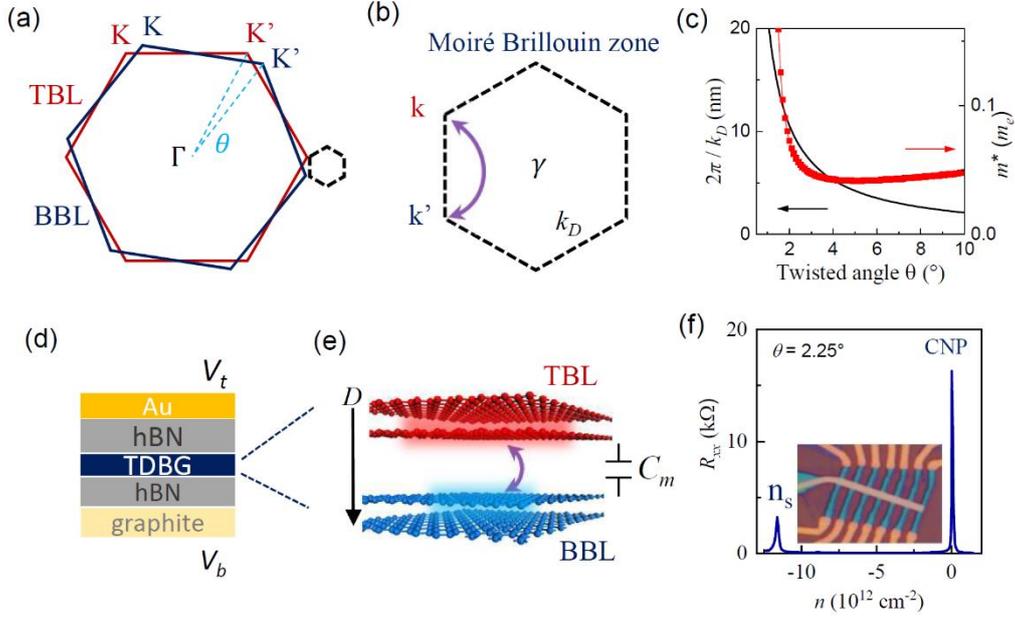

FIG. 1. A weakly coupled moiré system of twisted double bilayer graphene (TDBG). (**a**) shows the two Brillouin zones of top bilayer (TBL, red) and bottom bilayer (BBL, blue), with a twist angle ($\theta$). (**b**) is the zoomed-in of the resulted moiré Brillouin zone, with the moiré k (red) and k' (blue) from top and bottom layer, respectively. The curved arrow (purple) denotes the intervalley scatterings. (**c**) The calculated moiré periodicity (black) and the effective mass (red) as a function of twist angle in TDBG. (**d**) Sideview schematics of the stacked TDBG device structure. (**e**) Schematics of the interlayer capacitance ($C_m$) between the top bilayer (TBL) and bottom bilayer (BBL) in TDBG, with the curved arrow (purple) for the interlayer scatterings. (**f**) A typical transfer curve of $R_{xx}$ ($n$) at $T = 1.8$ K for device D1, and a twist angle of ~2.25º is obtained from carrier density at full fillings of moiré superlattice ($n_s$). The inset shows the Hall bar structure of D1.

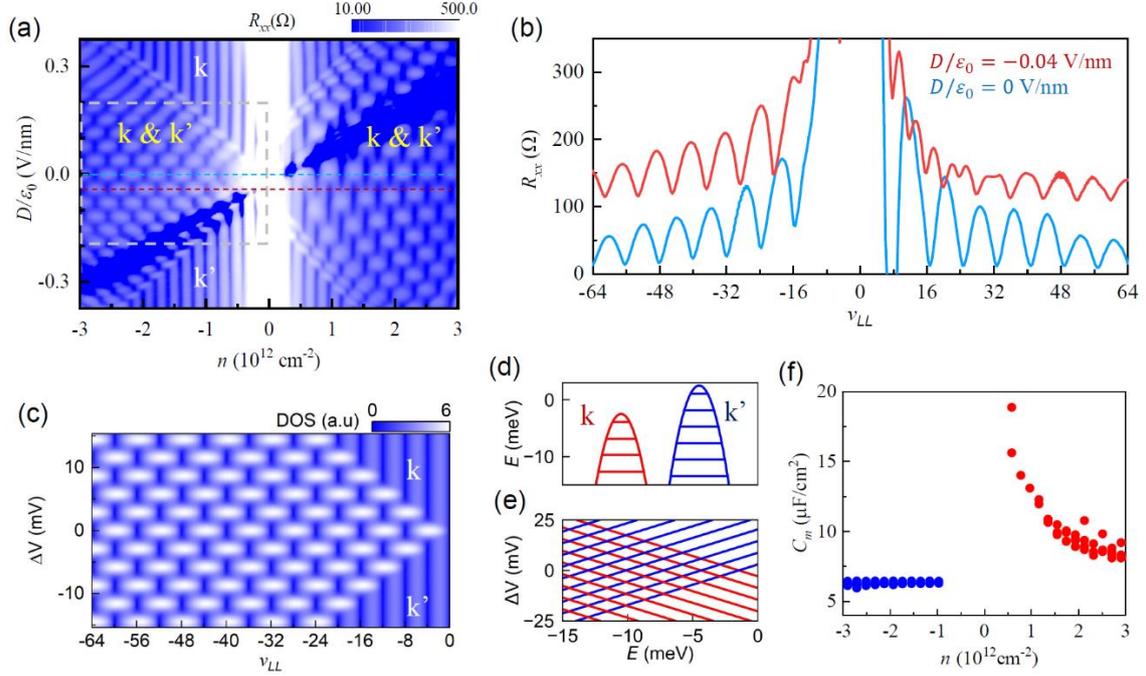

FIG. 2. Quantifying the interlayer coupling strength from the quantum oscillation spectra in a capacitance model. (a) A color mappings of $R_{xx}(v_{LL}, D/\varepsilon_0)$ at $T = 1.8$ K and $B = 2$ T. (b) plots of $R_{xx}$ at $D/\varepsilon_0 = 0$ and -0.04 V/nm, corresponding to the blue and red dotted lines in (d), respectively. The red line is offset by 100 Ω for clarity. (c) Calculated DOS as a function of Landau level (LL) filling factor $v_{LL}$ and interlayer potential difference $\Delta V$ at $B = 2$ T. (d) shows the schematic of the parabolic approximation for the minivalley $k$ (red) and $k'$ (blue) and the LLs structure in magnetic field for DOS calculation, and (e) is the evolution of the LLs spectrums of $k$ and $k'$ as a function of energy $E$ and the interlayer potential difference $\Delta V$ at $B = 2$ T. (f) The measured $C_m$ versus $n$, where $C_m$ are extracted from the intersection pattern in (a) according to the capacitance model.

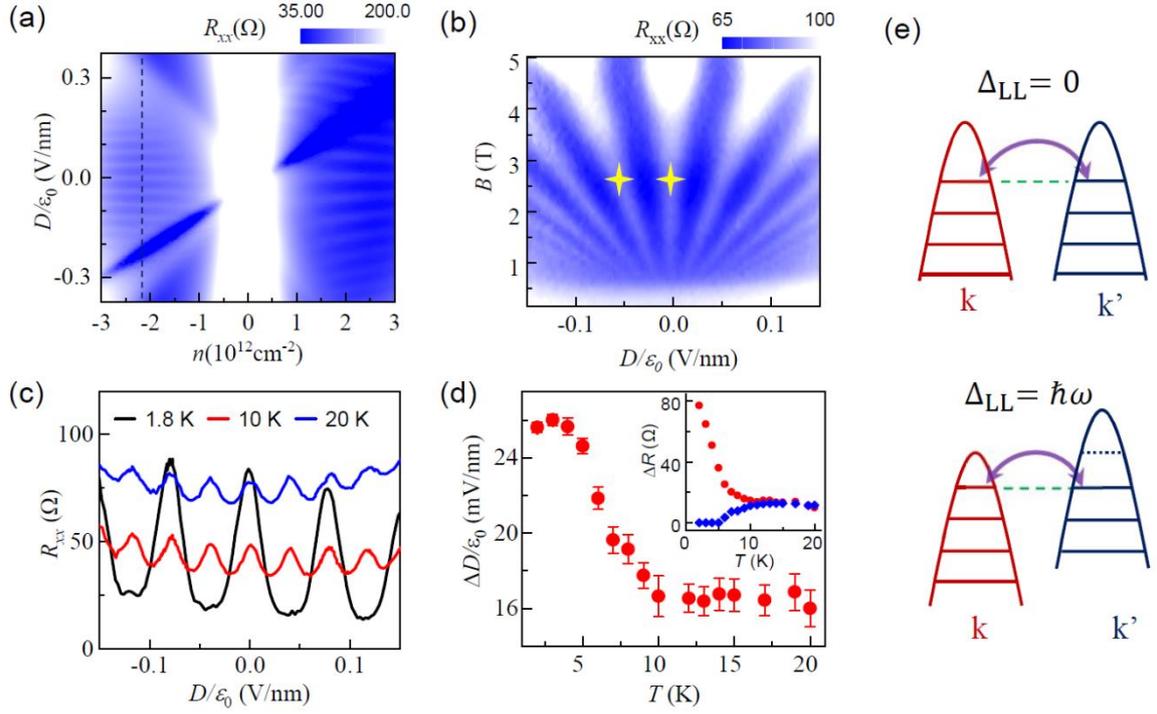

FIG. 3. The displacement field dependence of the carrier density insensitive quantum oscillations. (a) $R_{xx}$ as a function of $n$ and $D/\varepsilon_0$ at $B = 2$ T and $T = 20$ K. (b) $R_{xx}$ as a function of $D/\varepsilon_0$ and $B$ at $n = -2.18 \times 10^{12}$ /cm$^2$ and $T = 20$ K. (c) $R_{xx}$ versus $D/\varepsilon_0$ at $T = 1.8$ K, 10 K and 20 K. Here $B = 2$ T, $n = -2.18 \times 10^{12}$ /cm$^2$ corresponding to the blue dotted line in (a). (d) $\Delta D/\varepsilon_0$ versus $T$, where $\Delta D/\varepsilon_0$ is the FWHM of the resistance peak at $D/\varepsilon_0 = 0$ in (c). Inset: Amplitude of resistance peaks $\Delta R_{xx}$ versus $T$ at $D/\varepsilon_0 = 0$ (red) and $D/\varepsilon_0 = 0.043$ V/nm (blue). (e). Schematics of the enhanced intervalley scatterings at the resonant condition when LLs from k and k' are aligned, corresponding to the yellow stars in (b). The green dashed lines denote the fermi level.

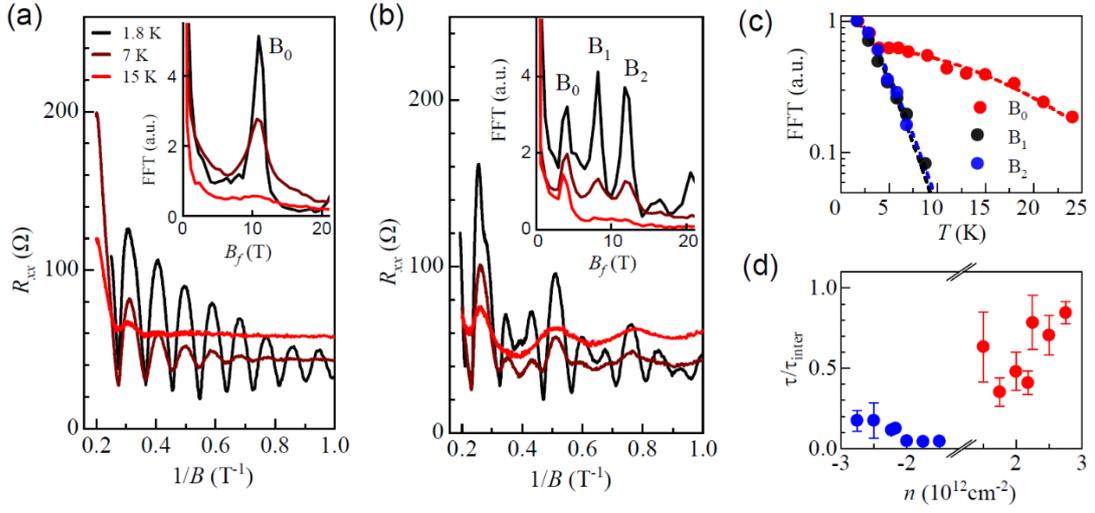

FIG. 4. Temperature dependence of the quantum oscillations. (a) and (b) are temperature dependent $R_{xx}(1/B)$ for $D/\varepsilon_0 = 0$ and -0.08 V/nm, respectively, where $n$ is fixed at $-2.16 \times 10^{12}$ /cm$^2$ and the inserts are FFT spectrums. (c) Normalized FFT amplitude verse $T$ for the mode $B_0$, $B_1$ and $B_2$ in (b). $B_0$ mode is fitted with $\exp(-\gamma T^2)$ decay, while $B_1$ and $B_2$ are fitted with L-K formula. (d) The ratio of $\tau/\tau_{inter}$ versus $n$ at $T = 18$ K, where $\tau/\tau_{inter}$ are extracted by fitting the high temperature magnetoresistance curve with the equation (3).